\documentclass{emulateapj}
\usepackage{graphicx}
\usepackage{amsmath}
\usepackage{rotating}
\usepackage{epsfig}
\newcommand{\degree}{\ensuremath{^\circ}}
\begin{document}
\title{Water deuterium fractionation in the inner regions of two
  solar type protostars
\footnote{Based on observations carried out with the IRAM Plateau de Bure Interferometer. IRAM is supported by INSU/CNRS (France), MPG (Germany) and IGN (Spain).}}
\author{V. Taquet$^{1,2}$,  A. L\'opez-Sepulcre $^{1}$,
  C. Ceccarelli$^{1}$, R. Neri $^{3}$, C. Kahane $^{1}$, A. Coutens
  $^{4,5}$, C. Vastel $^{4,5}$}
\altaffiltext{1}{UJF-Grenoble 1 / CNRS-INSU, Institut de
  Plan\'{e}tologie et d\textquoteright Astrophysique de Grenoble
  (IPAG) UMR 5274, Grenoble, F-38041, France}
\altaffiltext{2}{Current Address: Astrochemistry Laboratory and The
  Goddard Center for Astrobiology, Mailstop 691, NASA Goddard Space
  Flight Center, 8800 Greenbelt Road, Greenbelt, MD 20770, USA}
 \altaffiltext{3}{Institut de Radioastronomie Millim\'{e}trique
   (IRAM), 300 rue de la Piscine, 38406 Saint Martin d’H\'{e}res}
 \altaffiltext{4}{Universit\'{e} de Toulouse, UPS-OMP, IRAP, Toulouse,
   France}
\altaffiltext{5}{CNRS, IRAP, 9 Av. Colonel Roche, BP 44346, 31028
  Toulouse Cedex 4, France}

    \date{Received - ; accepted -}
\begin{abstract}

  The [HDO]/[H$_2$O] ratio is a crucial parameter for probing the
  history of water formation. So far, it has been measured for only
  three solar type protostars and yielded different results, possibly
  pointing to a substantially different history in their 
  formation. In the present work, we report new 
  interferometric observations of the HDO $4_{2,2}-4_{2,3}$ line
  for two solar type protostars, IRAS2A and IRAS4A,
  located in the NGC1333 region. In both sources, the detected HDO
  emission originates from a central compact unresolved
  region. Comparison with previously published interferometric
  observations of the H$_2^{18}$O $3_{1,3}-2_{2,0}$ line shows that
  the HDO and H$_2$O lines mostly come from the same region.  A
  non-LTE LVG analysis of the HDO and H$_2^{18}$O line emissions,
  combined with published observations, provides a [HDO]/[H$_2$O]
  ratio of 0.3 - 8 \% in IRAS2A and 0.5 - 3 \% in IRAS4A.
  First, the water fractionation is lower than that of other molecules
  such as formaldehyde and methanol in the same sources. Second, it is
  similar to that measured in the solar type protostar prototype,
  IRAS16293-2422, and, surprisingly enough, larger than that measured
  in NGC1333 IRAS4B. {The comparison of the measured values towards
  IRAS2A and IRAS4A with the predictions of our gas-grain model
  GRAINOBLE gives similar conclusions to those for IRAS 16293, arguing that
  these protostars {share} a similar chemical history, although they
  are located in different clouds.} 

 \end{abstract}


\section{Introduction}

The formation of solar type protostars is triggered by the
gravitational collapse of dense fragments of molecular clouds, the
so-called prestellar cores.  In molecular clouds and prestellar cores,
the low temperature and interstellar UV flux promote the formation of
icy mantles around the dust grains. These mantles are mainly composed
of H$_2$O, CO, CO$_2$, H$_2$CO, or CH$_3$OH
\citep[see][]{Boogert2004}.  The two last become abundant with the
freeze-out of CO.  {In addition, the CO freeze-out coupled with
  the cold conditions enhance the abundance of the deuterated
  molecules \citep[see][]{Ceccarelli2007, Caselli2008}. }

Several theoretical studies have shown that the molecular deuteration
is very sensitive to the physical conditions.  For instance, the
deuteration of gaseous species increases with the total density
$n_{\textrm{H}}$ and decreases with the temperature
\citep[see][]{Millar1989, Roberts2003}. Similarly, the deuteration of
icy species formed on the grain surfaces by H and D atom addition
reactions, like H$_2$O, H$_2$CO and CH$_3$OH, depends on the gaseous
atomic [D]/[H] ratio, which also increases with the density and the CO
freeze-out at low temperatures \citep{Cazaux2011, Taquet2012b,
  Taquet2013}. 

In theory, therefore, the deuteration of different mantle species can
be used to reconstruct the history of the ice formation and,
consequently, of the protostar \citep[e.g.][]{Taquet2013}.  In
practice, unfortunately, the direct measurement of the deuteration of
frozen species is not possible. {Observations of solid HDO
towards protostars only yielded upper limits \citep[with
HDO/H$_2$O $\lesssim$  few percent,][]{Dartois2003, Parise2003}.}
However, one can observe these species where the icy mantles
sublimate, for example in the hot corino regions. Since the timescale
needed to significantly alter the deuteration after the ice
sublimation is longer than the typical age of Class 0 protostars
\citep[$\sim 10^5$ versus $\sim 10^4$ yr,][]{Charnley1997, Andre2000},
the measured deuteration of the gaseous mantle species likely reflects
that in the ices prior to the sublimation. A comparison between the
measured and predicted deuteration in interstellar ices is, therefore, possible.

In \citet{Taquet2013}, we did a first study by comparing the
predictions of our gas-grain GRAINOBLE model \citep{Taquet2012a} with
the observations towards the protostar IRAS16293-2422 (hereafter,
IRAS16293). This source displays a very high deuterium fractionation
of formaldehyde and methanol \citep[with D/H ratios of 15 and 40 \%,
respectively, see][]{Loinard2001, Parise2002, Parise2004} and a lower
fractionation of water \citep[0.1 - 3 \%, see][]{Butner2007,
  Vastel2010, Coutens2012, Coutens2013, Persson2013}.  We concluded
that the lower fractionation of water with respect to that of
formaldehyde and methanol is likely due to a different epoch of
formation of the three species. Water is predicted to
be mainly produced during the molecular cloud phase, while most of
formaldehyde and methanol is formed during the colder and
denser prestellar phase. We carried out a similar study using the
measured deuteration of H$_2$O, H$_2$CO, and CH$_3$OH towards the
outflow shock L1157-B1 and concluded that this site had a similar
sequence for the formation of the ice, but in a less dense environment
\citep{Codella2012}.

Encouraged by these two studies, we want here to extend the analysis
to other solar type protostars with the goal to reconstruct the
formation history of their ices and compare it with the two previous
cases. Ultimately, a similar study in a large sample of solar type
protostars will provide us with a more complete picture of how the
environment influences the chemical composition of the ices and will
supply strong constraints to the theory.

Although the fractionation of formaldehyde and methanol has been
measured towards several solar type protostars \citep{Parise2002,
  Parise2004, Parise2005,Parise2006}, observational studies of
deuterated water are scarce. 
In NGC1333-IRAS4B, the non detection of the HDO line at 225.6 GHz
yields an upper limit to the [HDO]/[H$_2$O] ratio of $< 6 \times
10^{-4}$ \citep{Jorgensen2010}.
In NGC1333-IRAS2A, several HDO and H$_2$O lines have been observed
with single-dish telescopes. Using the \textit{Herschel Space
  Observatory}, \citet{Kristensen2010} observed a broad outflow
component {for several H$_2$O lines}, but could not accurately
estimate the water abundance in the warm compact region. In contrast,
\citet{Liu2011} derived the HDO abundance profile in the warm and cold
regions of the envelope. {However, single-dish telescopes also
  encompass the cold envelope, and the possible outflow
  component. Complementary interferometric observations, with
  arcsecond resolutions, are needed to resolve the emission coming
  from the hot corinos, where the ices are sublimated and where the
  deuteration likely reflects the ice pristine deuteration
  \citep{Jorgensen2010, Persson2013}. }

In this Letter, we present interferometric IRAM Plateau de Bure
observations of the HDO $4_{2,2}$-$4_{2,3}$ line at 143 GHz towards
NGC1333-IRAS2A (hereinafter IRAS2A) and NGC1333-IRAS4A (hereinafter
IRAS4A). These sources are located in the Perseus complex,
 in the NGC1333 cloud, whose distance is about 220 pc
\citep{Cernis1990}.  They were selected because they are the two
line brightest low-mass protostars after IRAS16293 due to their distance and their
luminosity and because interferometric observations of H$_2^{18}$O
have been recently obtained by \citet{Persson2012} towards them.
The [HDO]/[H$_2$O] ratio derived in the present work, combined with previous
observations of deuterated formaldehyde and methanol, will be compared
with the predictions of our gas-grain model GRAINOBLE
\citep{Taquet2013} to reconstruct the chemical history of these two
protostars.

\section{Observations and results} \label{obs}

The two low-mass Class 0 protostars IRAS2A and IRAS4A were observed
with the IRAM Plateau de Bure Interferometer on 2010 August 1, August
3 and 2011 March 10 and in the C and D configurations of the
array. Due to the proximity to each other, the two sources were
observed in the same track.  The $4_{2,2}$-$4_{2,3}$ HDO transition at
143.727 GHz and the 2 mm continuum emission have been obtained
simultaneously using the WIDEX correlator, with a 1.8 GHz bandwidth
centered at 143.5 GHz, and providing a spectral resolution of 1.95 MHz
(4 km s$^{-1}$).  Phase and amplitude were calibrated by performing
regular observations of the nearby point sources 3C454.3, 3C84, and
0333+321. The amplitude calibration uncertainty is estimated to be
$\sim 20$ \%.

The data calibration and imaging were performed using the CLIC, and
MAPPING packages of the GILDAS software\footnote{The GILDAS package is
  available at http://www.iram.fr/IRAMFR/GILDAS}. Continuum images
were produced by averaging line-free channels in the WIDEX correlator
before the Fourier transformation of the data. 
The coordinates of the sources, and the size of the synthesized beams
are reported in Table \ref{description}.  

\begin{table}[h]
\centering
\caption{Coordinates, synthesized
  beams, continuum fluxes, and sizes of the observed low-mass
  protostars.}
\begin{tabular}{l c c c c c}
\tableline
\tableline
Source & IRAS2A &  IRAS4A \\
\tableline
RA & 03:28:55.56 & 03:29:10.45 \\
Dec & 31:14:37.05 & 31:13:31.18 \\
Synthesized Beam (\arcsec) & 2.16 x 1.73 (25$\degree$) & 2.18 x 1.76
(25$\degree$) \\
Continuum flux (Jy) \tablenotemark{a}  & 0.13 & 1.08 \\
Source size (\arcsec) \tablenotemark{a}  & 1.75 x 1.69 & 2.13 x 1.72 \\
\tableline
\end{tabular}
\tablecomments{$^a$Continuum fluxes and sizes are obtained from elliptical
  Gaussian fits in the (u,v) plane (i.e., deconvolved full width at
  half-maximum (FWHM) size).} 
\label{description}
\end{table}

\begin{figure*}[h]
\centering
\includegraphics[width=50mm]{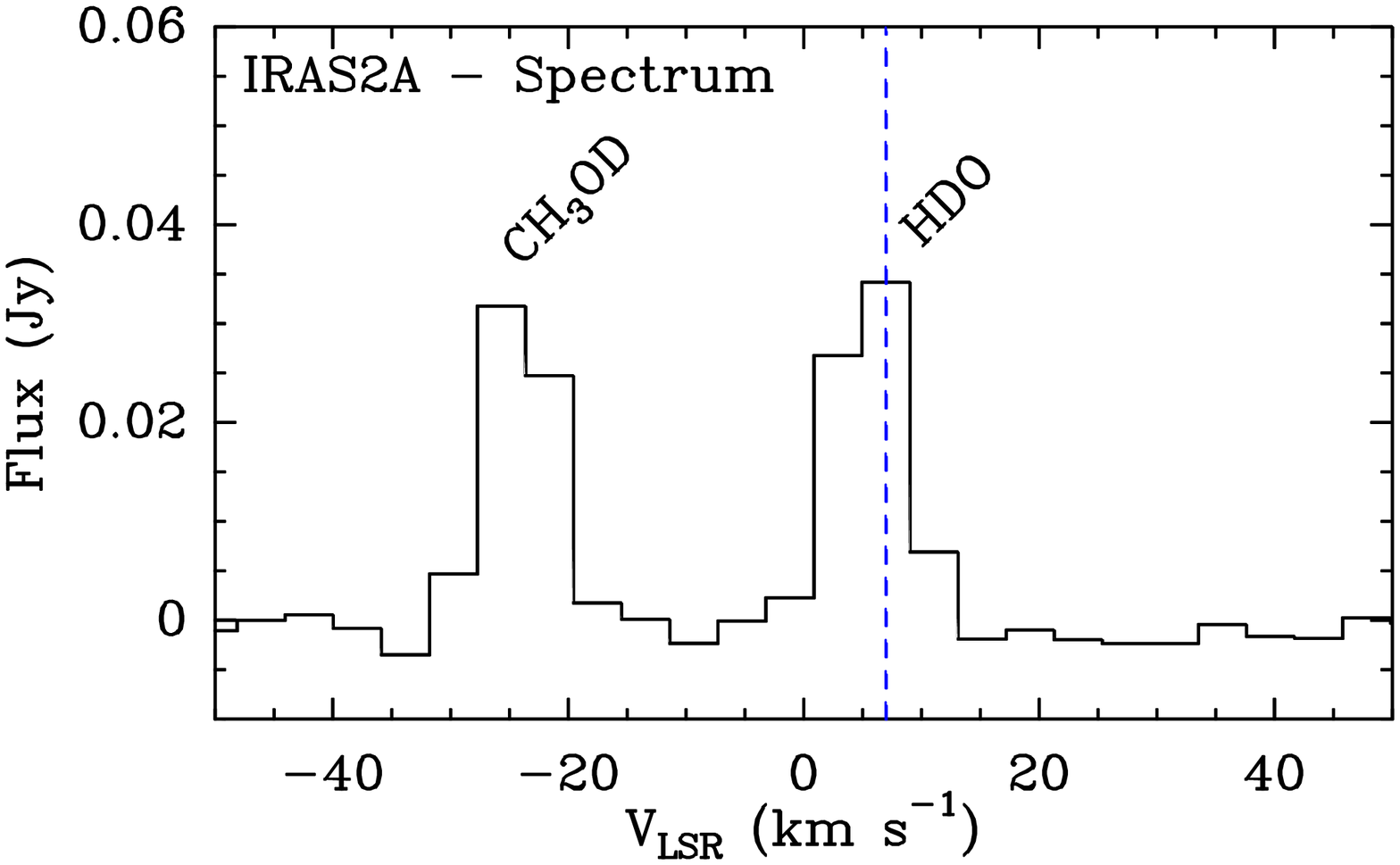}
\includegraphics[width=50mm]{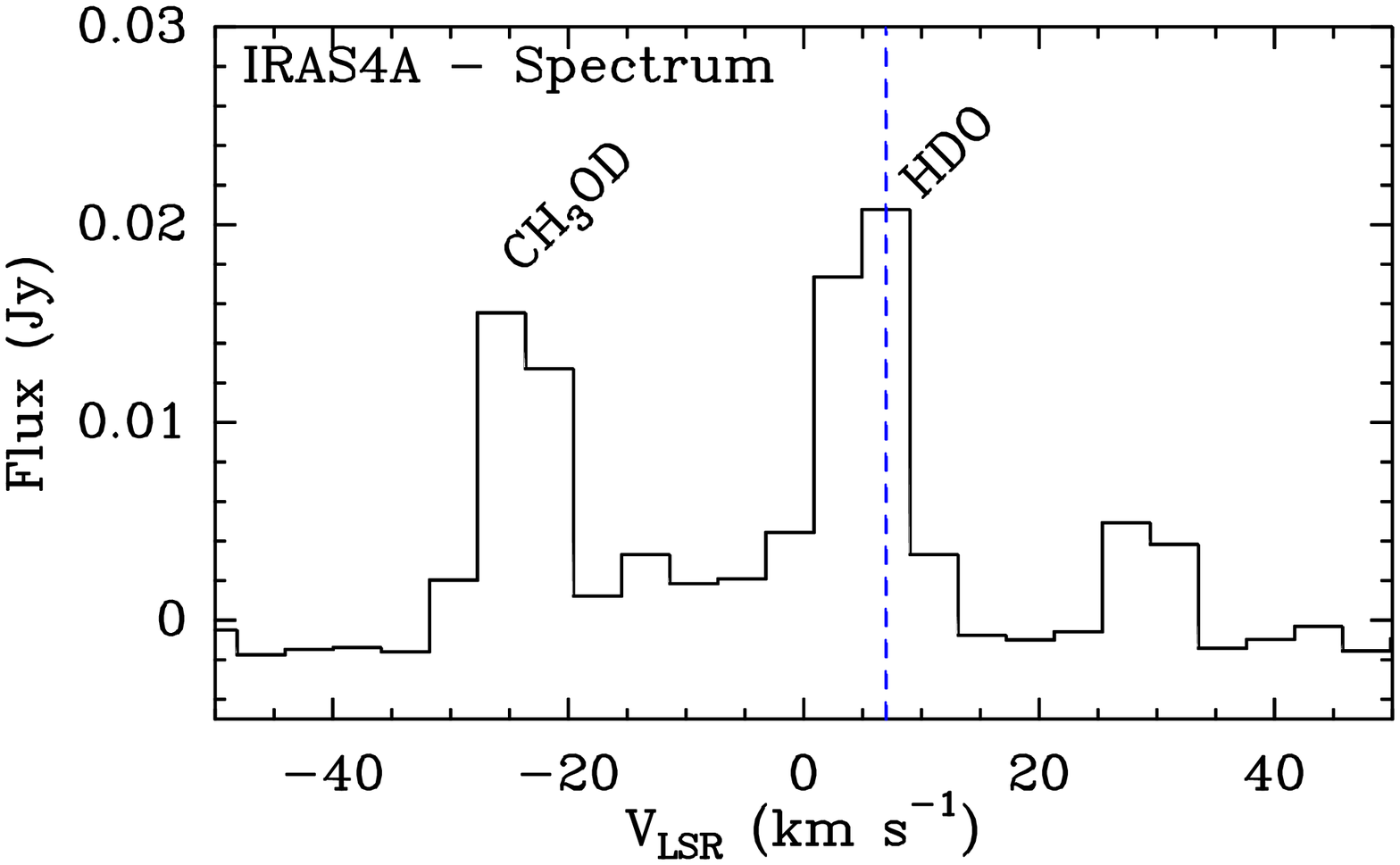} \\
\includegraphics[width=50mm]{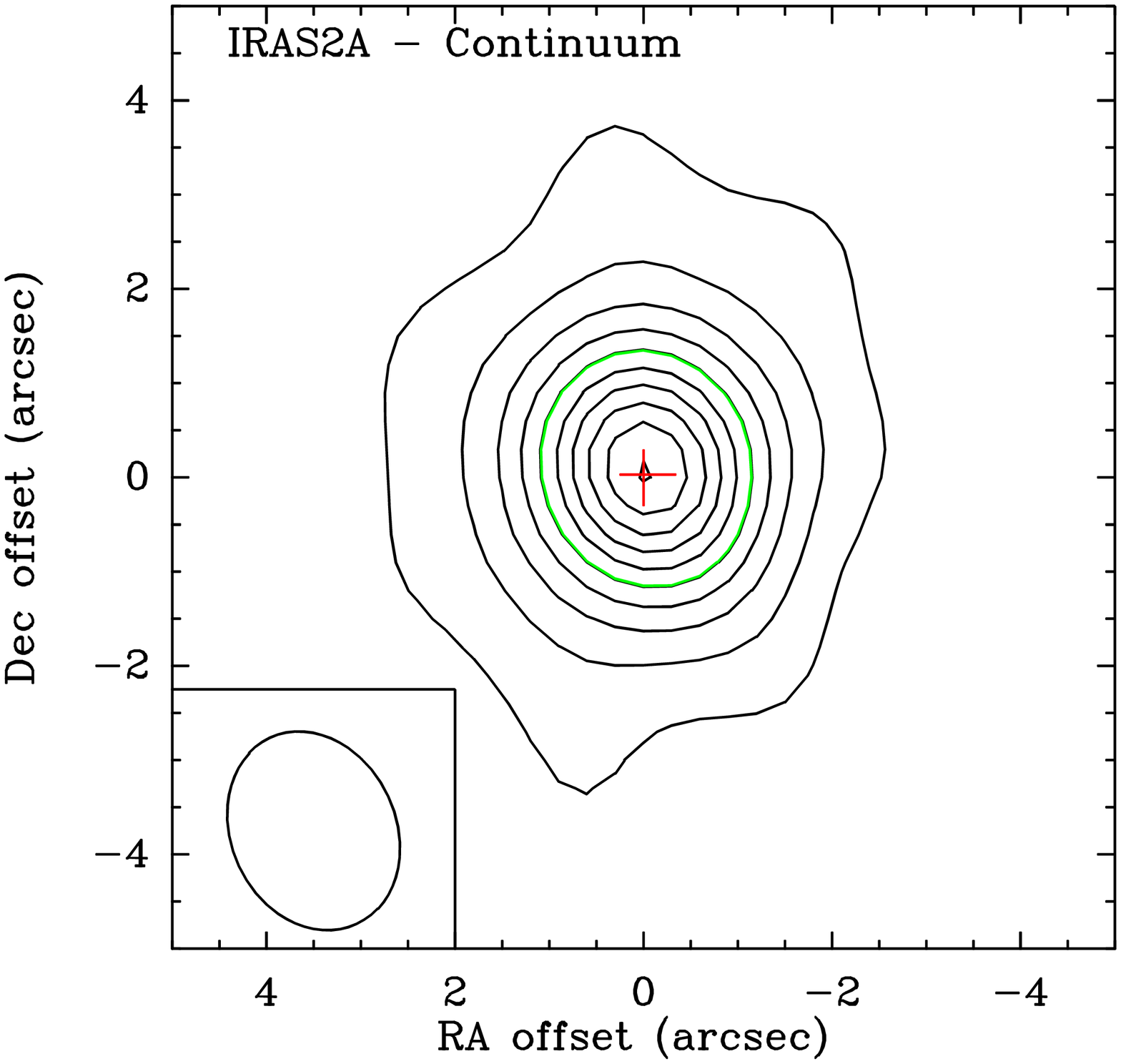}
\includegraphics[width=50mm]{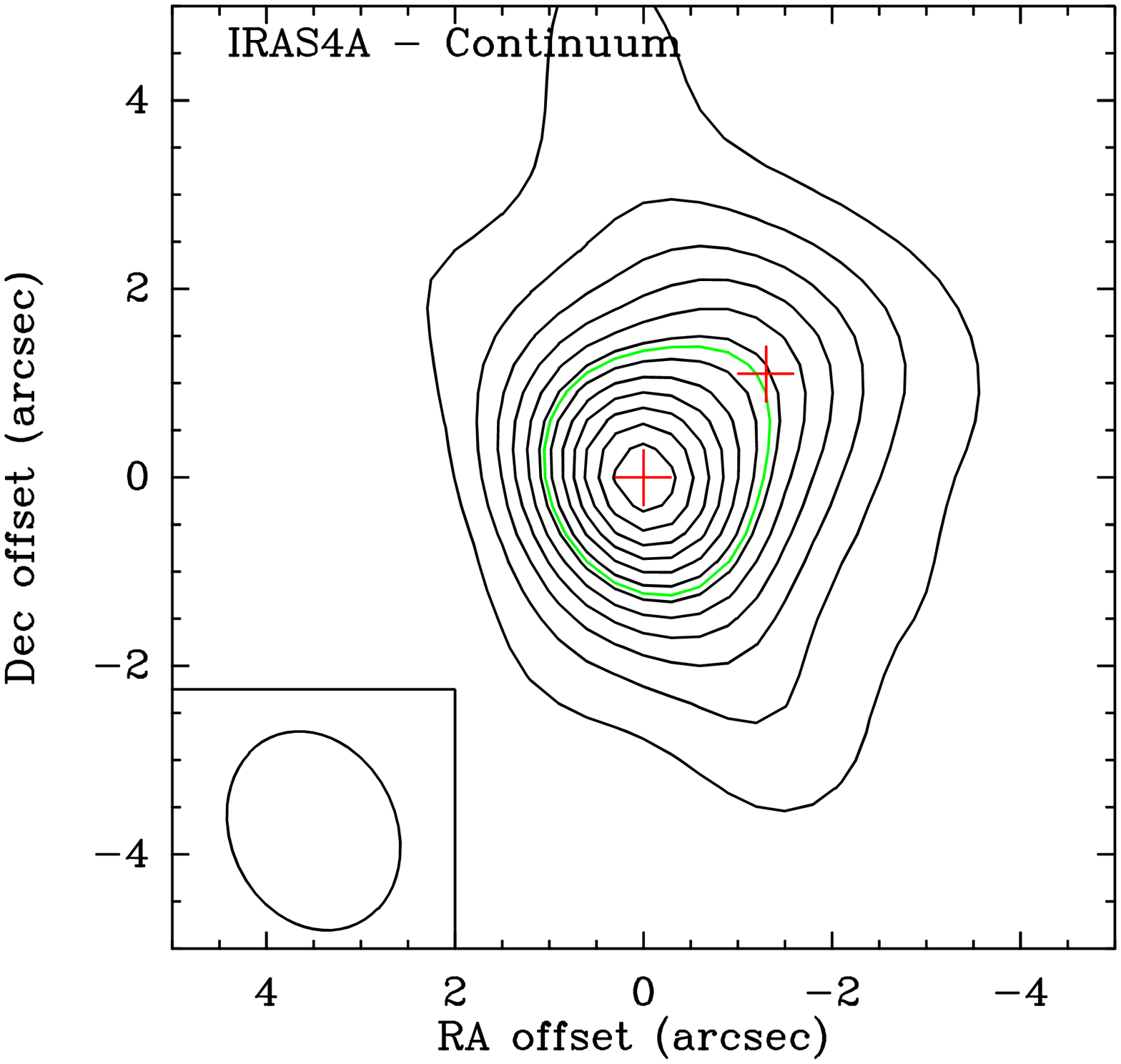} \\ 
\includegraphics[width=50mm]{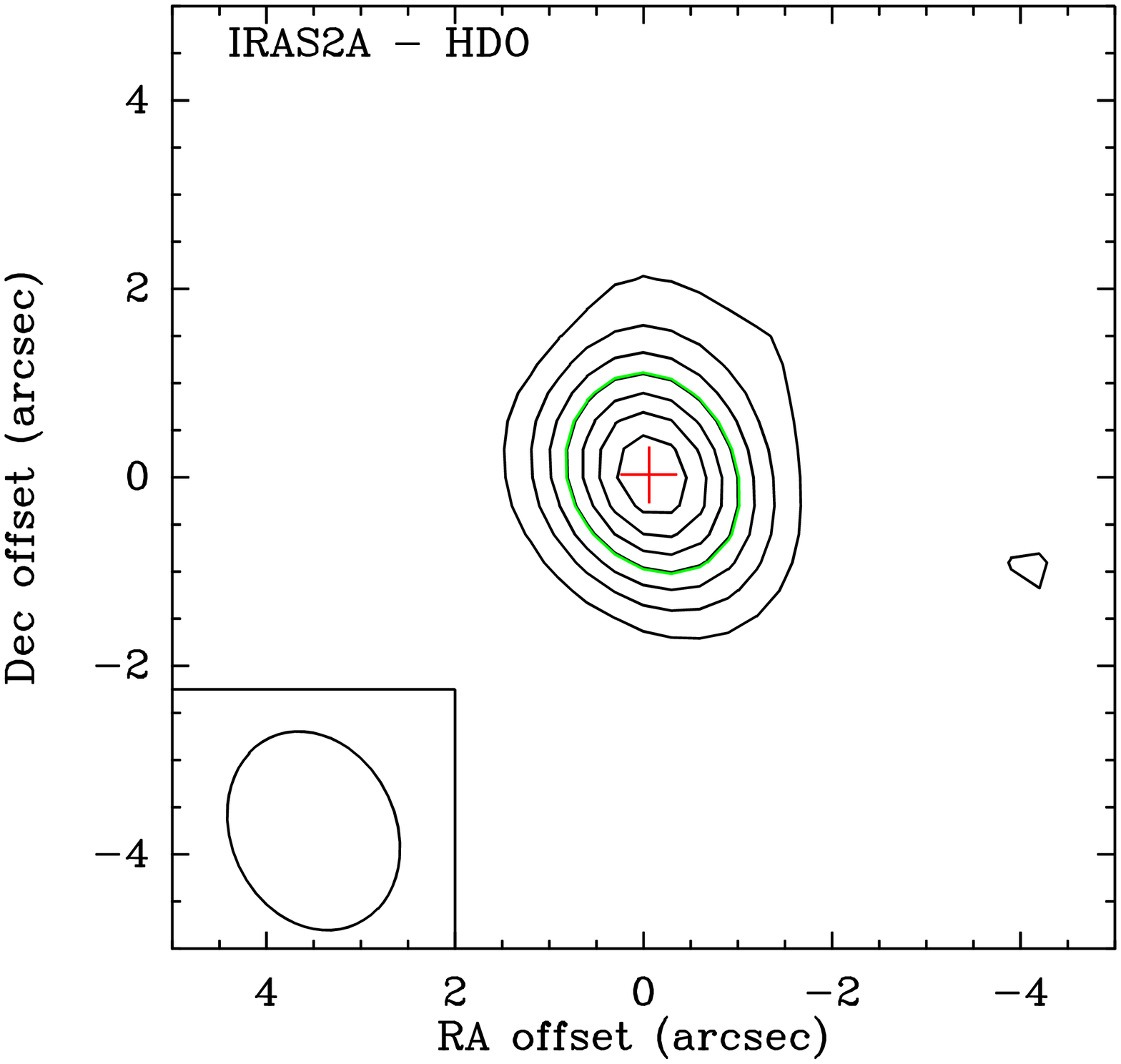} 
\includegraphics[width=50mm]{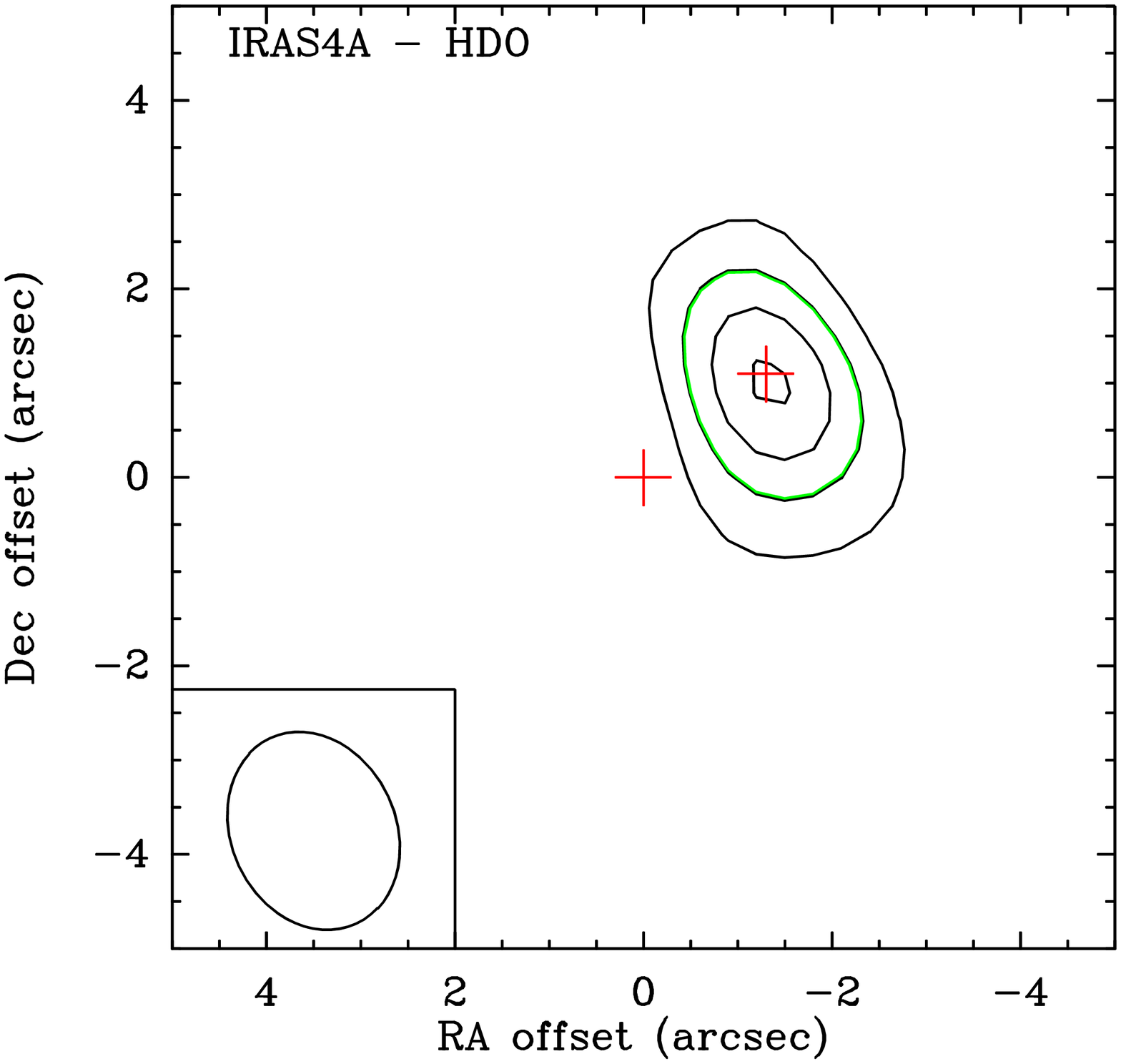}
\caption{Maps and spectra towards IRAS2A (left) and IRAS4A (right).
  Upper panels: HDO spectra integrated over the emission region of
  each source (middle panels). The velocity resolution is 4 km s$^{-1}$ 
  and the dashed blue lines mark the $V_{LSR}$ at 7 km s$^{-1}$.  
  Middle panels: Continuum maps at 143 GHz of IRAS2A (rms of 1.7
  mJy/beam, contour levels are in steps of 4 $\sigma$), and of IRAS4A
  (rms of 9.6 mJy/beam, contour levels are in steps of 4 $\sigma$).
  Green contours depict the deconvolved full width at half-maximum
  size.  
  Bottom panels: Maps of the HDO $4_{2,2}$-$4_{2,3}$ line towards
  IRAS2A (with a rms of 2.9 mJy/beam km s$^{-1}$, contour levels are
  in steps of 3 $\sigma$) and in IRAS4A (with a rms of 2.8 mJy/beam km
  s$^{-1}$, contour levels are in steps of 3 $\sigma$). Green contours
  depict the region showing a flux at half maximum of the line peak.
  The red crosses mark the source positions measured by
  \citet{Looney2000}. The bottom-left ellipses represent the beam
  sizes. }
\label{maps}
\end{figure*}

Figure \ref{maps} shows the maps of the continuum emission at 2 mm
of IRAS2A and IRAS4A obtained after natural weighted cleaning.
Parameters of the continuum emission (flux and deconvolved FWHM size),
obtained from elliptical Gaussian fits, are given in Table
\ref{description}.  For the two sources, since the FWHM size of the
continuum emission is very similar to the size of the synthesized
beam, the continuum emission is, therefore, not resolved. In
particular, IRAS4A is known to be a binary system with a
1.8\arcsec~separation \citep{Looney2000}, as depicted in Figure
\ref{maps}.  Although the continuum emission of IRAS4A is peaked at
the southeast (SE) position rather than at the northwest (NW)
position, we cannot resolve the two sources.

Figure \ref{maps} shows the maps of the integrated HDO
$4_{2,2}$-$4_{2,3}$ line towards the two sources obtained after
natural weighted cleaning.  For both sources, the FWHM size of the HDO
line is very similar to the size of the synthesized beam, as shown in
Figure \ref{maps}. The emission clearly originates in compact
regions, that are confined within the synthesized beam of the
telescope.  In particular, although the SE position of IRAS4A is the
brightest in the continuum, the HDO line emission comes from the NW
position.  The spectra of the HDO transition integrated within the
FWHM size are shown in Figure \ref{maps} assuming $V_{LSR} = 7$ km
s$^{-1}$ for the two sources. Table \ref{description2} gives the flux
and the brightness temperature of the HDO line transition integrated
inside the FWHM size.

\begin{table*}[h]
\centering
\footnotesize
\caption[Parameters of the HDO and H$_2^{18}$O lines observed towards
IRAS2A and IRAS4A.]{Parameters of the HDO and H$_2^{18}$O lines observed towards
IRAS2A and IRAS4A.} 
\begin{tabular}{l c c c c c c c c c c}
\tableline
\tableline
Transition & Frequency & $E_{up}$ & $A_{ij}$ & Flux & $(\int T_{\textrm{B}} dv)_{obs}$ & 
Beam & Telescope & $\Delta v$ & Ref. & \\ 
 & (GHz) & (K) & (s$^{-1}$) & (Jy km s$^{-1}$) & (K km s$^{-1}$) &
 (\arcsec) & & (km  s$^{-1}$) & & \\ 
\tableline
 & & & & \multicolumn{6}{c}{NGC1333 IRAS2A} \\
\cline{4-9} 
HDO $4_{2,2}$-$4_{2,3}$ & 143.727  & 319.2 & $3.5 \times 10^{-6}$ & 0.43 & $6.8 \pm
1.4$ & 2.2 x 1.8 & IRAM PdBi & 7 & 1  & \\ 
HDO $1_{1,0}$-$1_{1,1}$ & 80.578 & 46.8 & $1.3 \times 10^{-6}$ & 0.37
& $0.07 \pm 0.02$ & 31.2  & IRAM 30m & 3.9 & 2 & \\ 
HDO $2_{1,1}$-$2_{1,2}$ & 241.561 & 95.3 & $1.2 \times 10^{-5}$ & 2.2
& $0.43 \pm 0.05$ & 10.4 & IRAM 30m & 4.1 & 2 &  \\ 
HDO $3_{1,2}$-$2_{2,1}$ & 225.896 & 167.7 &$1.3 \times 10^{-5}$ &  2.6
& $0.50 \pm 0.03$ & 11.1 & IRAM 30m & 4.2 & 2 &  \\ 
H$_2^{18}$O $3_{1,3} - 2_{2,0}$ & 203.408 & 203.7 & $4.8 \times
10^{-6}$ & 0.98 & $46 \pm 9$ & 0.9 x 0.7 & IRAM PdBi & 4.0 & 3 & \\ 
\tableline
 & & & & \multicolumn{6}{c}{NGC1333 IRAS4A} \\
\cline{4-9} 
HDO $4_{2,2}$-$4_{2,3}$  & 143.727 & 319.2 & $3.5 \times 10^{-6}$ &
0.21 & $3.2 \pm 0.6$ & 2.2 x 1.8 & IRAM PdBi & 6 & 1 & \\ 
H$_2^{18}$O $3_{1,3} - 2_{2,0}$ & 203.408 & 203.7 & $4.8 \times 10^{-6}$ 
& 0.27 & $13 \pm 3$ & 0.9 x 0.7 & IRAM PdBi &  2.9 & 3 &  \\ 
\tableline
\end{tabular} \\
\tablerefs{1: This work, 2: \citet{Liu2011}, 3: \citet{Persson2012}.}
\tablecomments{The flux uncertainties include the calibration
  uncertainties, estimated to be $\sim 20$ \%.}
\label{description2}
\end{table*}
\normalsize

%
Comparison with maps of the H$_2^{18}$O $3_{1,3}$-$2_{2,0}$ line
transition by \citet{Persson2012} shows that, although our beam is
two times larger, most of the HDO and H$_2$O emissions originate in
the same region.   
\citet{Persson2012} estimated the FWHM size of the H$_2^{18}$O
emission, from a Gaussian fit in the $(u,v)$ plane, and they found that
most of the emission originates in a 0.8\arcsec~ellipse, similar to
their synthesized beam. Observational data derived by
\citet{Persson2012} are given in Table \ref{description2}. {The
  HDO lines observed in this work are broader, by a factor of 2, than
  the other HDO and H$_2^{18}$O lines, due to the low spectral
  resolution of our observations.} 
In IRAS2A, even if the bulk of the H$_2^{18}$O emission is associated
with the central warm envelope, an outflow component towards the southwest
is also observed \citep[see Fig. 2 of][]{Persson2012} whereas the HDO
line only originates in the compact region located within the
synthesized beam.
Our maps are, therefore, in good agreement with {previous single-dish
observations of H$_2^{16}$O and HDO} towards IRAS2A by
\citet{Kristensen2010} and \citet{Liu2011}, 
described in the Introduction, which show that H$_2$O {mostly} traces the
outflow whereas HDO only traces the central envelope.

In IRAS4A, both the HDO and H$_2^{18}$O emission come from the NW
position and the FWHM sizes are similar to the respective synthesized
beams, suggesting that the emission originates in the central region.
Therefore, it is meaningful to compare the flux of the HDO and
H$_2^{18}$O lines originating in the FWHM central regions. The fluxes
are given in Table \ref{description2} and are used in the next section
to estimate the [HDO]/[H$_2$O] ratio.

\section{Deuterium fractionation of water} \label{wat_deut}

\subsection{Method}

A single transition line of HDO and H$_2^{18}$O does not allow us to
derive an accurate estimate of the HDO and H$_2$O column densities
towards the two protostars and, therefore, of [HDO]/[H$_2$O].
In order to derive the physical conditions of the line emitting gas
and the relevant column densities, we compared the predictions from a
non-LTE LVG code \citep{Ceccarelli2003} with our observations and the
observations by \citet{Liu2011} of several HDO lines towards IRAS2A,
obtained with the IRAM 30m, JCMT, and APEX telescopes. We considered
the collisional coefficients from \citet{Daniel2011} for H$_2^{18}$O
and from \citet{Faure2012} for HDO. The Einstein coefficients are from
the Jet Propulsion Laboratory molecular database \citep{Pickett1998}.

We ran a grid of models covering a large parameter space in kinetic
temperature $T_{kin}$ (15 values from 70 to 220 K), $n_{\textrm{H}}$ (15
values from $1 \times10^6$ to $1\times10^9$ cm$^{-3}$), HDO column
density $N$(HDO) (15 values from $8\times10^{14}$ to $1\times10^{17}$
cm$^{-2}$), and source size $\theta_s$ (30 values from 0.1 to 200
arcsec). In addition, we considered three values for the ortho to para
ratio (opr) of H$_2$: $10^{-2}$ (namely all H$_2$ molecules are in the para state), 1, and 3
(thermal equilibrium value at $T > 50$ K).  To find
the best fit to the data, we excluded the 464 GHz line observed by
\citet{Liu2011} as it may be contaminated by the cold envelope
emission, given its low energy level (22 K).

\subsection{Results}

\paragraph{IRAS2A}
We ran the grid of models to reproduce the emission of the HDO lines
towards IRAS2A.  The H$_2$ opr has a low influence on the column
densities derived from the observations. Varying the H$_2$ opr
between 0.1 and 3 causes a small variation of the results, by no more
than 20 \%, namely within the uncertainties of the observations. In the
following, we consider an H$_2$ opr of 3.
The flux of all the HDO lines are well reproduced (reduced $\chi^2 <
1$) for $T_{kin} \sim$ 75-80 K, $\theta_s = 0.4$\arcsec,
and a wide range of $n_{\textrm{H}}$ between $6 \times 10^5$ and $2
\times 10^8$ cm$^{-3}$. The derived $N$(HDO), varies between $5 \times
10^{17}$ and $10^{19}$ cm$^{-2}$ and decreases with $n_{\textrm{H}}$.
To evaluate the H$_2$O column density $N$(H$_2$O) from the 203.4 GHz
transition, we considered three physical cases that reproduce the
emission of the HDO lines (see Table \ref{table:bestfit}).
The density used in Case 1 ($6 \times 10^5$ cm$^{-3}$) is
similar to the density used by \citet{Maret2004} for reproducing the
H$_2$CO emission with a non-LTE LVG analysis. The densities used in
Cases 2 and 3 are slightly lower than the density in the hot corino
region (where the temperature is higher than 100 K) of IRAS2A derived
by \citet{Jorgensen2002}.  Higher densities do not reproduce
the observed HDO emission (the reduced $\chi^2$ increases to
values much higher than 1).
Regardless of the density, the derived column density of H$_2^{18}$O
is equal to $6-7 \times 10^{16}$ cm$^{-2}$. Note that at $n_\textrm{H}
= 6 \times 10^5$ cm$^{-3}$, the line weakly masers \citep[see
also][]{Neufeld1991}. The column densities we obtain are slightly
higher, by a factor of two, than that derived by \citet{Persson2012}.
{The difference can, therefore, come from a combination of the
  LTE versus non-LTE population, gas temperature and line opacity (in
  our model, it is 1.4). The low temperature could indicate that
  the gas is thermally decoupled from the dust.} 
 $N$(H$_2^{16}$O) can then be derived by assuming an isotopic
 abundance ratio $^{16}$O/$^{18}$O of 560 \citep{Wilson1994} and an opr of
3 \citep[see][]{Emprechtinger2010, Emprechtinger2013}. Depending on
$n_{\textrm{H}}$, we derive an [HDO]/[H$_2$O] abundance ratio between
0.3 and 8 \% (see Table \ref{table:bestfit}).


\paragraph{IRAS4A}
For IRAS4A, no other HDO lines but the line observed in this work are
available. The flux of the HDO and H$_2^{18}$O lines are, therefore,
compared with the predictions obtained by using the same set of
physical conditions as for IRAS2A.  We also used another set of
parameters presenting a larger source size $\theta_s$ of 0.8\arcsec,
consistent with the upper limit given by \citet{Persson2012} for the
H$_2^{18}$O transition.  
{The increase in $\theta_s$ slightly decreases the column
  density of HDO and H$_2^{18}$O by approximately the same factor
(2-3),  giving similar results to those by \citet{Persson2012}. The
[HDO]/[H$_2$O] ratio, therefore, decreases by a factor of two, at
maximum.} For both sets of physical conditions, we predict an
[HDO]/[H$_2$O] abundance ratio between 0.5 and 3 \% (see Table
\ref{table:bestfit}).  

\begin{table}[htp]
\centering
\caption[Physical conditions and column densities of HDO and
  H$_2^{18}$O in IRAS2A and IRAS4A.]
{LVG best fit parameters for HDO and H$_2^{18}$O emissions.}
\begin{tabular}{l c c c}
\hline
\hline
Case & 1 & 2 & 3 \\
Density (cm$^{-3}$) & $6 \times 10^5$ & $2 \times 10^7$ & $1 \times 10^8$ \\
\hline
\multicolumn{4}{c}{IRAS2A ($T = 80$ K, $\theta_s = 0.4$ \arcsec)} \\
\cline{1-4}
$N$(HDO) (cm$^{-2}$) & $1 \times 10^{19}$& $1 \times 10^{18}$ & $6
\times 10^{17}$ \\
$\tau$(HDO $4_{2,2}$-$4_{2,3}$) & 23 & 2 & 1 \\
$N$(p-H$_2^{18}$O) (cm$^{-2}$) & $6 \times 10^{16}$ & $7 \times 10^{16}$
& $7 \times 10^{16}$ \\
$\tau$(p-H$_2^{18}$O) & -0.4 & 0.8 & 1.5 \\
$N$(H$_2$O) (cm$^{-2}$) & $1.3 \times 10^{20}$ & $1.6 \times 10^{20}$
& $1.6 \times 10^{20}$ \\
HDO/H$_2$O & $0.08$ & $0.006$ & $0.003$ \\
\hline
\multicolumn{4}{c}{IRAS4A} \\
\hline 
 & \multicolumn{3}{c}{$T = 80$ K, $\theta_s = 0.4$ \arcsec} \\
\cline{2-4}
$N$(HDO) (cm$^{-2}$) & $1.5 \times 10^{18}$& $3 \times 10^{17}$ & $2
\times 10^{17}$ \\
$\tau$(HDO $4_{2,2}$-$4_{2,3}$) & 3 & 0.8 & 0.4 \\
$N$(p-H$_2^{18}$O) (cm$^{-2}$) & $2.5 \times 10^{16}$ & $1.5 \times 10^{16}$
& $1.5 \times 10^{16}$ \\
$\tau$(p-H$_2^{18}$O) & -0.4 & 0.05 & 0.15 \\
$N$(H$_2$O) (cm$^{-2}$) & $5 \times 10^{19}$ & $3 \times 10^{19}$
& $3 \times 10^{19}$ \\
HDO/H$_2$O & $0.03$ & $0.01$ & $0.007$ \\
\cline{2-4}
 & \multicolumn{3}{c}{$T = 80$ K, $\theta_s = 0.8$ \arcsec} \\
\cline{2-4}
$N$(HDO) (cm$^{-2}$) & $5 \times 10^{17}$& $1 \times 10^{17}$ & $6
\times 10^{16}$ \\
$\tau$(HDO $4_{2,2}$-$4_{2,3}$) & 0.8 & 0.3 & 0.2 \\
$N$(p-H$_2^{18}$O) (cm$^{-2}$) & $1.5 \times 10^{16}$ & $6 \times 10^{15}$
& $5.5 \times 10^{15}$ \\
$\tau$(p-H$_2^{18}$O) & -0.2 & -0.02 & 0.08 \\
$N$(H$_2$O) (cm$^{-2}$) & $3 \times 10^{19}$ & $1.3 \times 10^{19}$
& $1.3 \times 10^{19}$ \\
HDO/H$_2$O & $0.016$ & $0.008$ & $0.005$ \\
\hline
\end{tabular} \\
\label{table:bestfit}
\end{table}

\section{Discussion and conclusions} \label{discussion}

The first result of this work is the relatively high water
deuteration, $\sim1$ \%, in IRAS2A and IRAS4A. In IRAS2A, this value is
compatible with the lower limit derived by \citet{Liu2011} in
the same source (see Introduction). In IRAS4A, this is the first
published estimate.

Second, as in IRAS16293 and L1157-B1, the water deuteration is lower,
by about one order of magnitude, than the deuteration of formaldehyde
and methanol in the same sources, previously measured by
\citet{Parise2006}.

Third, the water deuteration in IRAS2A and IRAS4A is very similar to
that measured in IRAS16293 by \citet{Coutens2012}, $\sim$3
\%{, but higher than the ratio derived by \citet{Persson2013}
  in the same source. 
The difference between the two results might come from the choice of
the method.  
\citet{Persson2013} derived the [HDO]/[H$_2$O] ratio
from few lines by assuming LTE population and optically thin
emission,  whereas the quoted column density implies a line opacity
$\sim 5$ and the 203 GHz line may maser (see above).  
On the contrary, \citet{Coutens2012} uses single-dish
observations which also encompass the cold envelope even though most
of the lines have $E_{up} >$ 50K, so that the contamination from the
outer cold envelope is accounted for. } 

The ratio is at least one order of magnitude larger than the value
measured in IRAS4B, $< 6 \times 10^{-4}$, by \citet{Jorgensen2010},
despite the fact that this source lies in the same molecular cloud,
NGC1333, as IRAS2A and IRAS4A and it is only $\sim 15$\arcsec~away
from IRAS4A \citep{Sandell1991}. To add to this oddity, the
deuteration of formaldehyde and methanol in IRAS4B is very similar to
that measured in IRAS2A and IRAS4. 

\begin{figure}[tb]
\centering
\includegraphics[width=88mm]{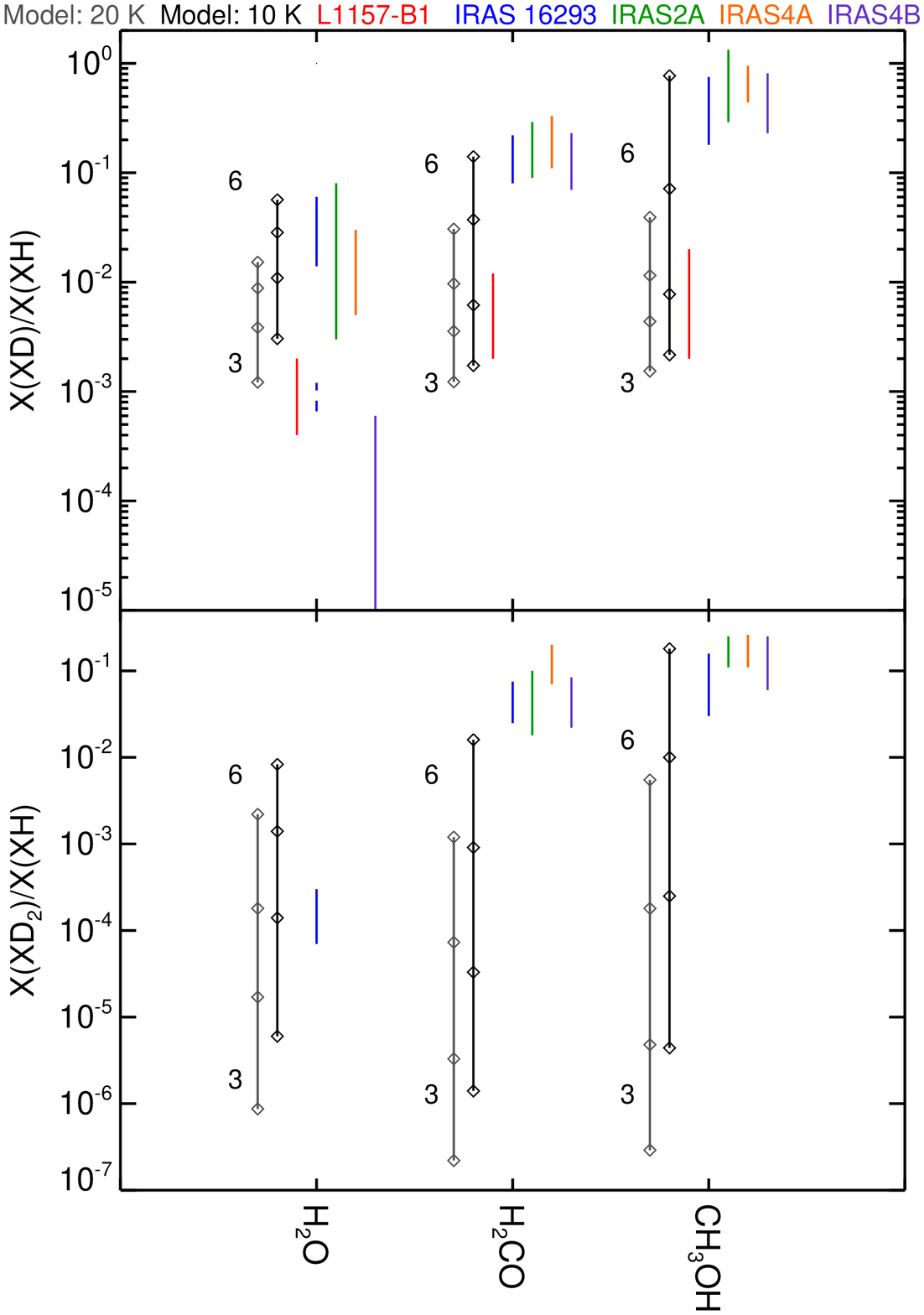}
\caption{Deuterium fractionation of water, formaldehyde, methanol of
  simply (top) and doubly (bottom) deuterated species.  From left to (with only
one exception
  right: predictions of GRAINOBLE model \citep{Taquet2013} at 20 K
  (grey) and 10 K (black), and observations 
  towards L1157-B1 \citep[red, from][]{Codella2012}, IRAS16293
  \citep[solid blue, from][and dashed blue from Persson et
  al. 2013]{Loinard2001, Parise2002, Parise2004, 
    Vastel2010, Coutens2012}, IRAS2A and IRAS4A \citep[green, from
  this work, for   water, and][for formaldehyde and
  methanol]{Parise2006}, and IRAS4B \citep[purple, from][]{Parise2006, 
    Jorgensen2010}. }
\label{obs_mod}
\end{figure}

Figure \ref{obs_mod} summarizes the situation, with a plot of the measured
deuteration of water, formaldehyde and methanol in the outflow shock
L1157-B1 and in the protostars IRAS 16293, IRAS2A, IRAS4A and IRAS4B.
{
In the same figure, we also show the
  theoretical predictions by the gas-grain model GRAINOBLE
  \citep{Taquet2013}. Briefly, the model follows the 
  multilayer formation of deuterated ices with a
  pseudo time-dependent approach. We report the icy [HDO]/[H$_2$O]
  ratio computed at $3 \times  10^5$ yr (the typical age of prestellar
  cores) for different constant densities and temperatures, $A_V = 10$
  mag. 
{The H$_2$ opr, which is difficult to constrain
  observationally, is one of the key parameters in setting the 
[HDO]/[H$_2$O] ratio \citep{Taquet2013}. Following the value derived
by \citet{Dislaire2012} towards IRAS 16293, we used a H$_2$ opr of
$10^{-3}$ . }
The comparison between the observations and the
theoretical predictions shows that the [HDO]/[H$_2$O] ratio measured
in IRAS2A and IRAS4A is reproduced for a large range of physical
conditions: $n_H \sim 10^3 - 10^5$ cm$^{-3}$ for $T=10$ K and $n_H
\sim 10^3 - 10^6$ cm$^{-3}$ for $T=20$ K.
%
On the contrary, our model cannot reproduce the 
[HDO]/[H$_2$O] value reported by \citet{Jorgensen2010} for densities
larger than $10^3$ cm$^{-3}$.  
{One possible explanation is that water ice has formed at a lower H$_2$
opr \citep[see][]{Taquet2013} or the model is missing some ingredients
regarding the deuterated ice formation.} 

As in our previous work, we note that the larger deuteration of
formaldehyde and methanol testifies to a formation of these species on
the grain surfaces at a later and higher density stage than water,
likely the prestellar core phase.  

Finally, NGC1333 is a very active star forming region undergoing the
destruction and alteration from various outflows of the first generation
stars {that might have initiated the formation of IRAS2A and IRAS4A}
\citep{Liseau1988, Warin1996, Lefloch1998, Knee2000} whereas the 
cloud containing IRAS16293 is relatively quiescent \citep{Mizuno1990}.
Nevertheless, the similar deuterium fractionation derived in IRAS2A,
IRAS4A and IRAS 16293 suggests that these protostars have followed a
similar chemical history even though they are located in very
different environments. {However, the [HDO]/[H$_2$O] ratio
  observed in IRAS4B by \citet{Jorgensen2010} and in IRAS 16293 by
  \citet{Persson2013} does not fit with this conclusion and remains
  puzzling.} 
 
\begin{acknowledgements}
This work has been supported by l\textquoteright Agence Nationale
pour la Recherche (ANR), France (project FORCOMS, contracts 
  ANR-08-BLAN-022).
\end{acknowledgements}

\end{document}